\documentstyle[11pt]{article}

\begin{document}
\begin{flushright}
\today
\end{flushright}
\vspace{0.5 cm}
\begin{center}
{\Large \bf IPM SCHOOL ON COSMOLOGY 1999}\\
\vspace{0.25 cm}
{\Large \bf LARGE SCALE STRUCTURE FORMATION}\\
\vspace{0.25 cm}
{\Large \bf JANUARY 23 -- FEBRUARY 4}\\
\vspace{0.25 cm}
{\Large \bf KISH UNIVERSITY, KISH ISLAND, IRAN}\\
\vspace{0.5 cm}
{\it \large Forough Nasseri $^{a, b, }$\footnote{e-mail: naseri@netware2.ipm.ac.ir}, 
Ali Nayeri $^{c, }$\footnote{e-mail: ali@iucaa.ernet.in}}\\
\vspace{3 mm}
$^{a}${\it \large Department of Physics, Sharif University of Technology, P.O.Box
11365--9161, Tehran, Iran}\\
$^{b}${\it \large Institute for Studies in Theoretical Physics and Mathematics,
P.O.Box 19395--5531, Tehran, Iran}\\
$^{c}${\it \large Inter-University Centre for Astronomy and Astrophysics,
Post Bag 4, Ganeshkhind, Pune -- 411 007, India}\\
\end{center}

The first IPM School on Cosmology 1999 on 
``Large Scale Structure Formation'',
sponsored by Kish Free Zone Organization, 
Kish University, ICTP (Italy),
UNESCO, Ministry of Culture and Higher Education, Meteorological 
Organization of Iran, Astronomical Society of Iran, was held at Kish 
University, from 23 January to 4 February 1999. The school brought
together
about 20 participants from 12 different countries 
besides 10 participants 
from Iran.\\
The theme of the School pertained to the recent advances made in
cosmology, 
like Big Bang cosmology, dark matter, origins of fluctuations,
super--symmetric Inflation, cosmological defects, string cosmology,
large scale structure formation, 
cosmic microwave background radiation (CMBR),
observational cosmology and high z--universe,
and time variation of $G$ and $\Lambda$.
These topics were explained by leading lecturers:\\
Alain Blanchard, Strasbourg, France\\
Robert Brandenberger, Brown university, USA\\
Joao Magueijo, IC, London, UK\\
Thanu Padmanabhan, IUCAA, India\\
Subir Sarkar, Oxford, UK\\
Matias Zaldarriaga, IAS, Princeton, USA.\\
Morning lectures were followed by exhaustive evening tutorials and
comments besides some seminars presented by:
N.~ Afshordi, S.~Arbabi, S.~Engineer, S.~Khakshornia, J.~Liske, M.~Moniez, 
F.~Nasseri, A.~Nayeri, S.~Pireaux, S.~Rahvar. \\
In addition, there were public lectures 
by Prof. T.~ Padmanabhan on ``Voyage to the Universe'' and by 
Prof. R.~H.~ Brandenberger on ``Structure Formation of the Universe''.
A public observation programme was done by S.~Arbabi, S.~Ghassemi and
M.~Moniez  at
Hour cottage.
In Sadaf girl's high-school a public lecture about astronomy was delivered 
by A.~ Nayeri.
The successful conduct of the school was mainly due to the active 
cooperation of the leading lecturers which is gratefully acknowledged.
The next IPM School on Cosmology will be held in 2002 on ``High
z-Universe 
and CMBR''.

\end{document}